# Effect of high pressure on multiferroic BiFeO$_3$

R. Haumont$^1$, P. Bouvier$^2$, A. Pashkin$^3$, K. Rabia$^3$, S. Frank$^3$, B. Dkhil$^4$

W. A. Crichton$^5$, C. A. Kuntscher$^3$, J. Kreisel$^{2, *}$

$^1$ ICMMO, Laboratoire de Physico-Chimie de l'Etat Solide (CNRS), Université Paris XI, 91405 Orsay, France

$^2$ Laboratoire des Matériaux et du Génie Physique, CNRS; Grenoble Institute of Technology, MINATEC, 38016 Grenoble, France

$^3$ Experimentalphysik 2, Universität Augsburg, 86135 Augsburg, Germany

$^4$ Laboratoire Structures, Propriétés et Modélisation des Solides, Ecole Centrale Paris, 92290 Châtenay-Malabry, France

$^5$ European Synchrotron Radiation Facility (ESRF), Grenoble, France

$^*$ Author to whom correspondence should be addressed

abstract
## Abstract

We report experimental evidence for pressure instabilities in the model multiferroic BiFeO$_3$ and namely reveal two structural phase transitions around 3 GPa and 10 GPa by using diffraction and far-infrared spectroscopy from synchrotron sources. The intermediate phase from 3 to 9 GPa crystallizes in a monoclinic space group, with octahedra tilts and small cation displacements. When the pressure is increased further the cation displacements (and thus the polar character) of BiFeO$_3$ is suppressed above 10 GPa. The non-polar orthorhombic *Pnma* structure observed above 10 GPa is in agreement with recent *theoretical* ab-initio prediction, while the intermediate monoclinic phase has not been predicted theoretically.

# I. Introduction

So-called magnetoelectric multiferroics, which exhibit both magnetic order and ferroelectricity in the same phase, have recently attracted a renewed fundamental interest. In particular, the prospect of using coupling between magnetic and ferroelectric degrees of freedom opens new perspectives in magnetic and/or ferroelectric storage media.[1-4]

Bismuth ferrite $BiFeO_3$ (BFO) is commonly considered to be a model system for multiferroics[5], especially for $ABO_3$ perovskites where the ferroelectricity is driven by an *A*-cation with $6s^2$ lone pair electrons. The perovskite BFO is one of the very few robust multiferroics with ferroelectric and antiferromagnetic order well above room temperature: In bulk material BFO has an antiferromagnet Néel temperature $T_N$ of ~380 °C and a ferroelectric Curie temperature $T_C$ of ~830 °C.[6, 7]

In recent years BFO has attracted an increasing interest following a report of an enhanced ferroelectric polarization of 60 µC/cm$^2$ in epitaxial thin films.[8] Early values reported[9] for the polarization of bulk BFO were rather small (8.9 µC/cm$^2$). The large polarization in thin films was initially ascribed to the effect of heteroepitaxial strain and thus to a change in lattice parameters with respect to the bulk.[8] However, subsequent[10] first-principles calculations have shown that the electric polarization in BFO is not affected significantly by the presence of epitaxial strain but is rather intrinsic to BFO.[11] This picture has been recently supported[12] by measurements on high-quality BFO ceramics[11] and crystals[13] for which a polarization of 40 µC/cm$^2$ - close to theoretical predictions - has been observed. Finally, the possibility of ferroelectric domain engineering and the report of both ferroelastic and ferroelectric switching processes suggest that a further modification and optimization of ferroelectric properties and the magnetoelectric coupling in epitaxial BFO films is in reach[3, 14-16], just as for ferroelectric thin films [17-20].

Much progress in understanding multiferroics has been achieved in recent years by investigating the effect of temperature, of an electric (magnetic) field and/or changes in chemical composition. Very little is known about the effect of high-pressure on magnetoelectric multiferroics and this despite the parameter pressure having played in the past a crucial role in the understanding of classic[21-28] and complex[29-35] ferroelectrics, or even more generally in transition metal oxides[36]. The external high-pressure parameter can be considered as a "cleaner" variable, compared to other parameters since it acts only on interatomic distances. In particular, the energetic order between different phases in perovskite materials can be notably modified by applying external pressure.

The room-temperature structure of $BiFeO_3$ is a highly rhombohedrally distorted perovskite with space group $R3c$.[37, 38] With respect to the cubic $Pm\bar{3}m$ structure the rhombohedral structure is obtained by an anti-phase tilt of the adjacent $FeO_6$ octahedra and a displacement of the $Fe^{3+}$ and $Bi^{3+}$ cations from their centrosymmetric positions along $[111]_{pc}$. As a consequence of this BFO presents, in addition to the magnetic order parameter, further ferroelectric and ferroelastic order parameters and a complex interplay between these different instabilities should be expected. A recent Raman scattering study has suggested[39] that $BiFeO_3$ undergoes two phase transitions below 10 GPa but the symmetry of the high-pressure phases (and thus the involved transition mechanism) remain to be discovered. Further to this experimental work, theoretical *ab-initio* based calculations have predicted a single pressure-induced structural transition from the initial rhombohedral $R3c$ structure to an orthorhombic *Pnma* ($GdFeO_3$-type) structure around 13 GPa.[40] Finally, very recent experimental[41-45] and theoretical investigations[46] discuss the occurrence of magnetic and electric phase transitions at 50 GPa in BFO, but phase transitions below 50 GPa are not observed. The fact that the latter authors do not observe a structural phase transition in BFO below 50 GPa is surprising when we recall that ferroelectric instabilities are known to be very sensitive to pressure (all pressure-investigated ferroelectric perovskites show at least one structural phase transition below 15 GPa).

The aim of our study is to verify experimentally the occurrence of the pressure-induced phase transition sequence in BFO below 20 GPa, to determine the symmetry of the two new phases and to reveal the phase transition mechanism. For this, we have undertaken a pressure-dependent X-ray diffraction and far-infrared spectroscopy study by using synchrotron radiation. We note that high-pressure infrared studies of phonon modes are rare in the literature, mainly because of the experimental difficulties when compared with Raman scattering. To the best of our knowledge, the present work is the first systematic study of the phonon behaviour in ferroelectrics under high pressure by means of infrared reflection spectroscopy.

## II. Experimental

### A. Sample preparation

The investigated single crystals of BiFeO$_3$ were grown using a Fe$_2$O$_3$/Bi$_2$O$_3$ flux in a platinum crucible. Red-translucent crystals with a shape of thin platelets have been isolated and Laue back-scattering indicates a [001]$_{pc}$ orientation of the platelet (pseudo-cubic setting). BFO powders were prepared by conventional solid-state reaction using high-purity (better than 99.9%) bismuth oxide Bi$_2$O$_3$ and iron oxide Fe$_2$O$_3$ as starting compounds. After mixing in stoichiometric proportions, powders were calcined at $T_\text{f}$ = 820°C for 3h. More synthesis details can be found in ref [12, 47].

### B. Synchrotron X-ray diffraction

Repeated high-pressure synchrotron X-ray diffraction experiments were performed at the European Synchrotron Radiation Facility (ESRF) on the ID09A high-pressure beam line. The powder sample was loaded in a diamond anvil cell (DAC) with diamond tips of diameter 350 µm and with hydrogen as a pressure-transmitting medium to assure good hydrostatic conditions up to the highest investigated pressure of 37 GPa. The pressure was measured using the ruby fluorescence method.[48] X-ray diffraction patterns were collected in an angle-resolved geometry on an image plate MAR345 detector with a focused monochromatic beam. The sample to detector distance, the wavelength $\lambda = 0.4110$ Å and the detector inclination angles were calibrated using a silicon standard. After removal of spurious peaks coming from the diamond cell, the two-dimensional diffraction images were analyzed using the ESRF Fit2D software [49], yielding intensity *vs.* 2θ diffraction pattern. XRD pattern after pressure release are identical to the initial attesting the reversibility of pressure-induced changes up to 37 GPa. The powder diffraction data were analyzed by full Rietveld refinements using the FullProf [50] software.

## C. Synchrotron far-infrared micro-spectroscopy

Pressure-dependent far-infrared reflectivity measurements at room temperature were carried out at the infrared beamline of the synchrotron radiation source ANKA in Karlsruhe (D) using a Bruker IFS 66v/S Fourier transform infrared spectrometer. A diamond anvil cell equipped with type-IIA diamonds suitable for infrared measurements was used to generate pressures up to 10 GPa. To focus the infrared beam onto the small sample in the pressure cell, a Bruker IR Scope II infrared microscope with a 15x magnification objective was used.

The measurement of the infrared reflectivity has been performed on the surface of as-grown $BiFeO_3$ crystals. A small piece of sample (about 80 µm × 80 µm × 40 µm) was placed in the hole (150 µm diameter) of a steel gasket. With this crystal size and the corresponding diffraction limit, we were able to measure reliably the frequency range above 200 cm$^{-1}$. Finely ground CsI powder was added as a quasi-hydrostatic pressure-transmitting medium. The ruby luminescence method was used for the pressure determination.[48]

Reflectivity spectra were measured at the interface between sample and diamond. Spectra taken at the inner diamond-air interface of the empty cell served as the reference for normalization of the sample spectra. The absolute reflectivity at the sample-diamond interface, denoted as $R_{s-d}$, was calculated according to $R_{s-d}(\omega) = R_{dia} \times I_s(\omega)/I_d(\omega)$, where $I_s(\omega)$ denotes the intensity spectrum reflected from the sample-diamond interface and $I_d(\omega)$ the reference spectrum of the diamond-air interface. The reference reflectivity of the diamond-air interface $R_{dia} = 0.167$ was calculated using the Fresnel equation with the refractive index of diamond, $n_{dia} = 2.38$, assumed to be independent over the range in pressure investigated. This is justified because $n_{dia}$ is known to change only very little with pressure.[51] Variations in synchrotron source intensity were taken into account by applying additional normalization procedures. Two experimental runs on different crystals ensured the reproducibility between datasets. The orientation of the samples in the pressure cell allowed us to probe the response of the phonon modes polarized normal to the direction of spontaneous polarization, similar to Ref. [52].

### III. Results

#### A. Synchrotron X-ray diffraction

We have performed a structural analysis of BiFeO$_3$ under high-pressure up to 37 GPa. Figure 1 displays the diffraction patterns obtained for three selected pressures. With increasing pressure, we observe significant changes in the multiplicity and intensity of the Bragg peaks that are indicative of two structural phase transitions at $p_{c1, XRD}$ = 3.6 GPa and $p_{c2, XRD}$ = 10 GPa, these values are close to earlier reported values determined by Raman scattering [39].

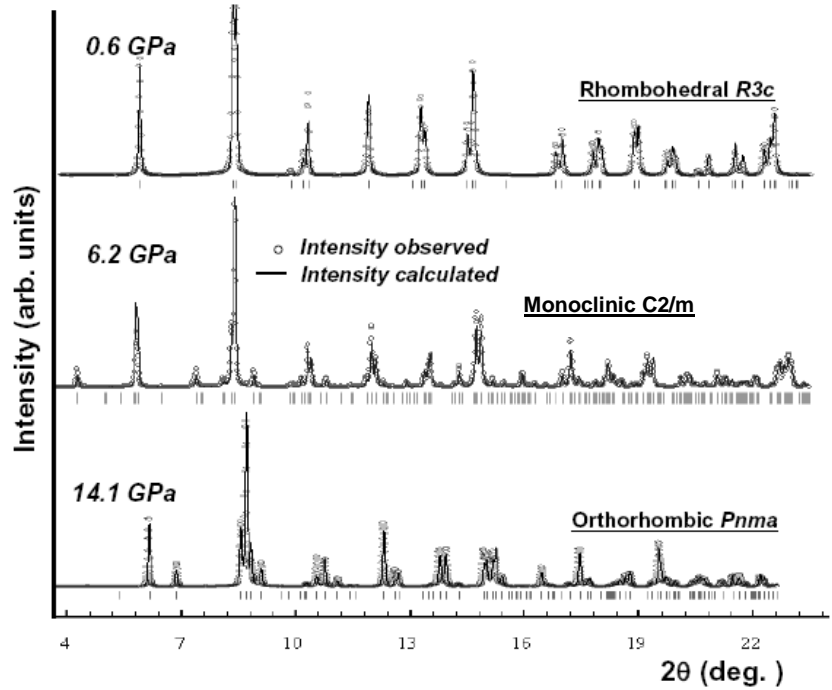

**Figure 1**
Rietveld refinement diffraction patterns of BiFeO$_3$ at three selected pressures (0.6 GPa, 6.2 GPa and 14.1 GPa) representing rhombohedral, monoclinic and orthorhombic symmetries respectively.

In order to have a better understanding of the phase transition mechanism we have performed Rietveld refinements analysis at 0.6, 6.2 and 14.1 GPa corresponding to the three different phases as pure phases. As expected, the low pressure phase at 0.6 GPa pressure is well described by a *R3c* rhombohedral symmetry (R). The unit cell parameters measured at ambient condition in the DAC are found to be equal to $a$ = 5.578(2) Å and $c$ = 13.865(3) Å (hexagonal setting) in good agreement with the crystal structure of BFO reported in the literature[38, 53]. Beside this, a small amount of 0.9% in volume of a Bi$_{25}$FeO$_{40}$ impurity phase is observed. The parameters obtained by fitting the *P–V* up to 3.6 GPa with a third-order Birch–Murnaghan equation of state (EoS)[54] using a pseudo-cubic cell (Z=1) are $V_0$ = 62.29(2) Å$^3$, $K_T$

= 111(6) GPa and $K'$ = 3.7(3). This value is slightly lower than the value obtained from ab-initio calculations.[40] We note that the ratio $c/a\sqrt{6}$ decreases progressively with pressure and reaches a value of 1.0048 at $p_{c1,\ XRD}$= 3.6 GPa. Thus at this transition pressure the volume approaches a metrically cubic cell.

Above 3.6 GPa, the phase transition to a new phase is evidenced from the appearance of numerous weak reflections in the diffraction pattern (Fig. 1). Initially, a metric orthorhombic cell is found using CRYSFIRE[55] unit-cell determination software. However, the symmetry assignment using CHECKCELL[56] failed in this orthorhombic cell. An approximate solution was obtained in the *P*1 space group using FOX[57] with 12 FeO$_6$ octahedra and 12 Bi atoms and a dynamic occupancy correction. We then performed a symmetry search on this approximate solution using ENDEAVOUR[58] and found a *C*2/*m* monoclinic symmetry. A final Rietveld refinement lead to reasonable agreement factors (chi²= 3.61, $R_{bragg}$ = 18.4%) for the diffraction pattern at 6.2 GPa. This monoclinic phase *C2/m* is non-ferroelectric and is characterized by a strong distortion due to FeO$_6$ octahedra tilting, implying a large unit cell ($Z$ = 12) associated with a large monoclinic angle $\beta$ = 108.24°. Table 1 summarises the refinement details. Although the monoclinic phase provides the best refinement parameters, we remind that perovskite structures are known for their structural subtleties and namely the history of PbZr$_{1-x}$Ti$_x$O$_3$ (PZT) has shown that new phases have to be confronted with a possible phase coexistence and/or micro/nano-twinning.[59-62] Within this context, we note that one can create a monoclinic cell (or metric orthorhombic cell) by addition of two twinned domains of adjacent R and O structures (or two O structures) or a phase coexistence. Because of the important number of refinement parameters, our data does not allow a reliable refinement of such scenario, but we note that Arnold et al.[63] have very recently reported a rhombohedral-orthorhombic phase coexistence for BFO at high temperature.

|     | C2/m           | (6.2 GPa)      |                |                     |
| --- | -------------- | -------------- | -------------- | ------------------- |
|     | $a_O$ = 17.5218(3) Å | $b_M$ = 7.7244(3) Å | $c_M$ = 5.4711(3) Å | $\beta_M$ = 108.24(18)° |
|     | x              | y              | z              | B (Å²)              |
| Bi  | 0.2525(8)      | 0              | 0.5035(7)      | 3.62                |
| Bi  | 0.4171(6)      | 0              | 0.2320(4)      | 3.64                |
| Bi  | 0.9135(8)      | 0              | 0.1999(9)      | 3.45                |
| Fe  | 0.25           | 0.25           | 0              | 4.12                |
| Fe  | 0.4123(6)      | 0.2399(8)      | 0.6912(8)      | 3.86                |
| O1  | 0.3949(29)     | 0              | 0.5243(34)     | 2.84                |
| O2  | 0              | 0.7296(29)     | 0              | 2.68                |
| O3  | 0.3169(34)     | 0.1842(36)     | 0.7534(32)     | 3.21                |
| O4  | 0.2500(21)     | 0.1032(29)     | 0.9061(33)     | 3.54                |
| O5  | 0              | 0.7073(28)     | 0.5            | 2.96                |
| O6  | 0.9136(32)     | 0              | 0.6444(29)     | 3.02                |
| O7  | 0.2994(35)     | 0              | 0.1164(28)     | 2.81 :              |
|     | Chi² = 3.61    | $R_{Bragg}$ = 18.4 % |          |                     |

**Table 1.** Result of the Rietveld X-ray-diffraction refinement at 6.2 GPa (C2/*m* space group).

At higher pressure, i.e. at 14.1 GPa, the refinement is more straightforward and the orthorhombic (O) *Pnma* space group converges to a satisfactory fit (chi²= 3.48, $R_{bragg}$ = 13.8%, see table 2). Note that the *Pnma* structure is commonly observed in perovskites, namely in the related Rare Earth orthoferrites $R$FeO$_3$ (R=Rare Earth) and has also been proposed as the high-temperature structure of BFO.[63] The parameters obtained by fitting the $P$–$V$ between 12 and 37 GPa with a third-order Birch–Murnaghan [54] equation of state (EoS) in pseudo cubic cell (Z=1) are $V_0$ = 56.41(2) Å³, $K_T$ = 238(5) GPa and $K'$ = 2.2(5).

|     | Pnma           | (14.1 GPa)     |                |         |
| --- | -------------- | -------------- | -------------- | ------- |
|     | $a_O$ = 5.4951(1) Å | $b_O$ = 7.6496(3) Å | $c_O$ = 5.3273(2) Å |  |
|     | x              | y              | z              | B (Å²)  |
| Bi  | 0.0465(4)      | 0.25           | 0.9896(7)      | 3.50    |
| Fe  | 0              | 0              | 0.50           | 4.06    |
| O1  | 0.5095(36)     | 0.25           | 0.0848(34)     | 2.81    |
| O2  | 0.5604(42)     | 0.0289(29)     | 0.2065(32)     | 2.16    |
|     | Chi² = 3.48    | $R_{Bragg}$ = 13.8 % |          |         |

**Table 2.** Result of the Rietveld X-ray-diffraction refinement at 14.1 GPa (Pnma).

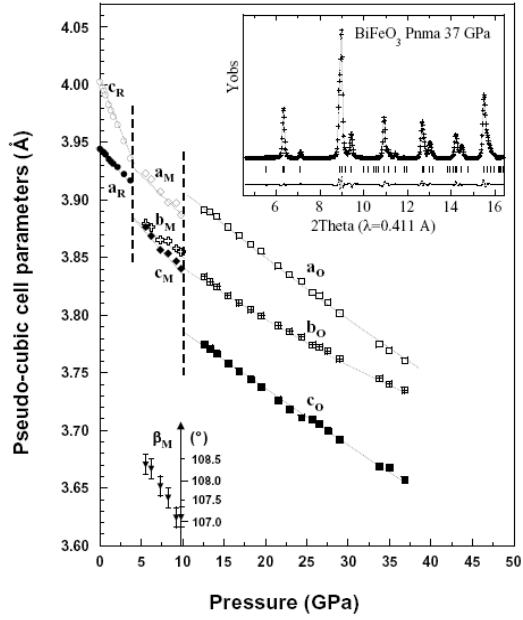

**Figure 2**
Pressure-dependent evolution of pseudo-cubic cell parameters of BiFeO$_3$. The inset shows a part of the diffraction pattern obtained at the highest investigated pressure of 37 GPa, attesting that the orthorhombic *Pnma* structure is maintained even up to this pressure.

Before discussing the pressure-induced monoclinic and orthorhombic phases in more detail, we first analyse the pressure-dependence of the lattice parameters, which is in itself instructive. The refined lattice parameters and their metric relationship with the pseudo-cubic parameters are shown in Table 3. Figure 2 displays the pressure-dependence of R, M and O cell parameters, expressed as a pseudo-cubic cell. In the low-pressure region, Figure 2 shows that the rhombohedral phase is very sensitive to pressure as we observe a significant decrease of both $a_R$ and $c_R$. A sharp decrease with pressure, commensurate with a large compressibility of 1.8 10$^{-2}$ Å.GPa$^{-1}$, is observed in P-$c_R$. This value is ~2.5 times more than $a_R$ and tends to merge with $a_R$ when approaching $p_{c1}$. This first phase transition at $p_{c1}$ from *R3c* to *C2/m* corresponds to a change in cation displacements (parallel to anti-parallel) and a change in the oxygen tilting system from ($a^-a^-a^-$) to ($a^-b^-c^0$) in Glazer's notation[64]. Moreover, within the monoclinic region the lattice parameters decrease with increasing pressure leading to a compressibility that is similar to that of $a_R$ in the rhombohedral region. Furthermore, the $\beta_M$ angle decreases slowly from 108.24° (6.2 GPa) to 107.13° (9.8 GPa), but does not reach 90° before the M → O transition.

| R3c | C2/m | Pnma |
|---|---|---|
| 0.6 GPa | 6.2 GPa | 14.1 GPa |
| $a_R$ = 5.5713(3) Å<br>$c_R$ = 13.8255(2) Å | $a_M$ = 17.5218(3) Å<br>$b_M$ = 7.7244(3) Å<br>$c_M$ = 5.4711(3) Å<br>$\beta_M$ = 108.24(18)° | $a_O$ = 5.4951(1) Å<br>$b_O$ = 7.6496(3) Å<br>$c_O$ = 5.3273(2) Å |
| $a_R = a_{pc} + b_{pc}$<br>$c_R = 2a_{pc} + 2b_{pc} + 2c_{pc}$ | $a_M = 2a_{pc} + 4c_{pc}$<br>$b_M = 2 b_{pc}$<br>$c_M = a_{pc} - c_{pc}$ | $a_O = a_{pc} + c_{pc}$<br>$b_O = 2 b_{pc}$<br>$c_O = a_{pc} - c_{pc}$ |
| $a_R = \sqrt{2}\ a_{pc}$<br>$c_R = 2\sqrt{3}\ a_{pc}$ | $a_M = 2\sqrt{5}\ a_{pc}$<br>$b_M = 2\ a_{pc}$<br>$c_M = \sqrt{2}\ a_{pc}$ | $a_O = \sqrt{2}\ a_{pc}$<br>$b_O = 2\ a_{pc}$<br>$c_O = \sqrt{2}\ a_{pc}$ |
| Z = 6 | Z = 12 | Z = 4 |

**Table 3**
Experimental BiFeO$_3$ cell parameters at 0.6 GPa, 6.2GPa and 14.1 GPa, with vectorial relations between rhombohedral, orthorhombic, monoclinic and pseudo-cubic cells parameters.

The rhombohedral structure of BFO is antiferrodistorsive with a FeO$_6$-tilting of 13.8° at ambient conditions [38] leading to $a^-a^-a^-$ tilts (Glazer notation [64]). Further structural refinements performed show that the tilting angle decreases as the pressure increases as it becomes equal to 7.9° and 5.3° at 1.4 GPa and 2.6 GPa respectively. We recall that pressure usually favours oxygen tilting rotation [21, 34] while some exceptions have been reported [65]. In case of BiFeO$_3$ the pressure-induced reduction of both the initially important oxygen tilting angle and the cation displacements allows to relax the elastic energy. However, while the ferroelectricity disappears above $p_{c1}$, the tilt angle does not reach zero and instead persists above $p_{c1}$ similar to earlier reported results on lead-based perovskites [34, 66].

A further increase of the pressure induces a phase transition at $p_{c2}$ from the monoclinic phase to the *Pnma* orthorhombic phase, which is non-polar, but with a distortion due to $a^+b^-b^-$ octahedra tilts. The cell parameters of this high-pressure orthorhombic phase present an almost linear pressure-dependent evolution. In particular, $a_O$ and $c_O$ have a similar coefficient of ~ 5.5 $10^{-3}$ Å.GPa$^{-1}$ while that of $b_O$ is weaker with a value of 4.2 $10^{-3}$ Å.GPa$^{-1}$. Interestingly, the extrapolation of the cell parameters leads to an intersection around 47 GPa; a pressure where $a_O = b_O \neq c_O$ leading to a tetragonal metric symmetry when assuming a second order transition. We also note that this extrapolated phase transition pressure may correspond to earlier reports on magnetic and electronic phase transitions of BFO in the same pressure range [43-45] which have been proposed to lead to a cubic *Pm-3m* symmetry.[67] Further investigations are needed to reveal the true structural behaviour and potential phase transitions in the high-pressure regime.

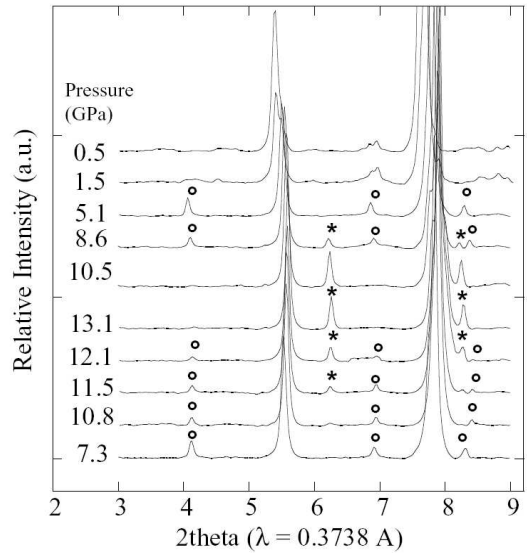

**Figure 3**

Low-angle X-ray diffraction patterns of BiFeO$_3$ at several selected pressures. The bottom 4 patterns correspond to increasing pressure, while the correspond top 6 patterns patterns correspond to pressure release, illustrating both the reversibility of the pressure-induced changes and the regimes of phase coexistence.

We further note that the discontinuous changes in cell parameters at $p_{c1}$ and $p_{c2}$ indicate that both $R \rightarrow M$ and $M \rightarrow O$ phase transitions are of first order. This observation is further supported by the presence of a phase coexistence of the two adjacent phases as evidenced for the second phase transition in the diffraction pattern in Figure 3. A group-subgroup relation does not exist between either the rhombohedral $R3c$ and monoclinic $C2/m$ or between monoclinic $C2/m$ and orthorhombic $Pnma$ space groups, which is consistent with first order phase transitions. Note that the $C2/m$ phase is peculiar as the bismuth-cations are "artificially" set on two different sites whereas the symmetry imposes that all four Bi-cation sites in this structure would be symmetrically equivalent. This situation is similar to previously observed high temperature behaviour[68] and probably arises from the electronic lone pair associated to the $s$-orbital of the bismuth. We return to this point below.

It is instructive to set our above results into the larger context of bismuth-based Bi$M$O$_3$ perovskites, since the unusually distorted monoclinic phases seem to be a common feature, most probably conditioned by the existence of the electronic lone pair arising from the $s$-orbital. For instance, the structure of BiMnO$_3$ (even though if its exact symmetry is still debated in the literature [69-72]), presents a highly distorted monoclinic ($C2$, $C2/m$ or $Cm$) symmetry with a ~ 9.5 Å, b ~ 5.6 Å, c~ 9.86 Å and β ~ 108.6° [71, 72]), and a complex sequence of phase transitions under temperature. Moreover, a monoclinic phase also describes the structure of BiCrO$_3$ that crystallizes in the $C2$ space group [73] or BiScO$_3$ with a $C2/c$ phase with a ~ 9.89 Å, b ~ 5.82 Å, c ~ 10.04 Å and β ~ 108.3° [74]. It is interesting that all the above Bi-based perovskites share three common features *(i)* they are thermodynamically stabilized and synthesized under pressure; *(ii)* the unit cells present a large distortion; and in particular,

*(iii)* $\beta_M$ is always close to 108.3°. Based on the above considerations, it is useful to understand the occurrence of the pressure-induced monoclinic phase in $BiFeO_3$. Our work underlines the fact that the monoclinic phases observed in the Bi-based perovskite are metastable at ambient conditions but may be stabilized under pressure. This finding might also explain the monoclinic structure observed in $BiFeO_3$ epitaxial thin film [8]. Figures 4a and 4b display a scheme of the projection of this monoclinic phase in the ($a_M$ ; $c_M$) and ($a_M$ ; $b_M$) respectively wherein we propose a geometrical configuration for Bi-lone pair respecting crystallographic and chemical considerations that are very close to that suggested for $BiScO_3$ [16] based on neutron diffraction and electronic microscopy:

- The two-fold axis 2 along $b_M$ and mirror *m* (located at y = ¼ and y = ¾) perpendicular to this axis impose that lone pairs are along the $b_M$ axis, at coordinates y = 0 and y = 0.5.

- Such configuration is also compatible with an *a* mirror generated by the combination of a two-foldaxis 2 and a mirror *m* represented by dotted line on figure 4; and also with the $2_1$ screw axis parallel to $b_M$, and generated at z = ¼.

- If we consider the absence of the lone pair in the Figure 4 then it can be seen that the two Bi-sites become symmetry-equivalent allowing a description of the structure in an orthorhombic setting, which would in turn allow visualizing the M → O phase transitions in a different way.

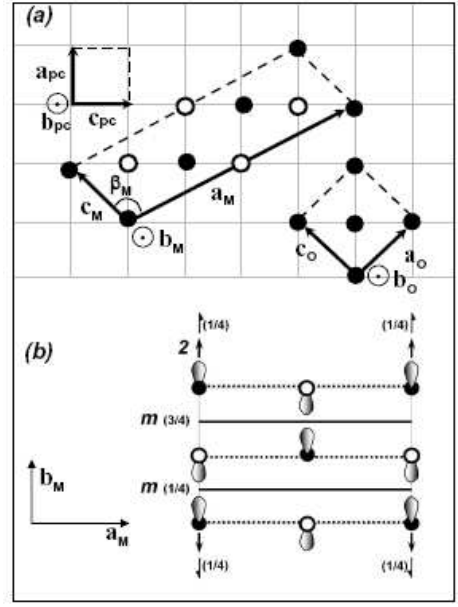

**Figure 4**
*(a)* Build of the low pressure monoclinic ($a_M$, $b_M$, $c_M$, $\beta_M$) and the high pressure orthorhombic ($a_O$, $b_O$, $c_O$) cell vectors in the pseudo-cubic ($a_{pc}$, $b_{pc}$, $c_{pc}$) vectorial base. Bismuth atoms, taken at is origin, are represented by black and white open circles, according the direction of its lone electron pairs (symbolised by grey lobes) along $b_M$.
*(b)* In-plane ($a_M$, $b_M$) projection of Bi positions, in respect with two-fold *2* axis and *m* mirror in the monoclinic *C2/m* symmetry.

## B. Synchrotron far-infrared micro-spectroscopy

Figure 5 presents the far-infrared reflectivity spectra of BiFeO$_3$ at room-temperature for three selected pressures; the spectra are offset along the vertical axis for clarity. Following the analysis of the infrared and terahertz spectra in Ref. [75], we applied the generalized-oscillator model to our spectra with the factorized form of the complex dielectric function:

$$\varepsilon(\omega) = \varepsilon_\infty \prod_{j=1}^{n} \frac{\omega_{LO_j}^2 - \omega^2 + i\omega\gamma_{LO_j}}{\omega_{TO_j}^2 - \omega^2 + i\omega\gamma_{TO_j}} \qquad (1)$$

where $\omega_{TO\,j}$ and $\omega_{LO\,j}$ denote the transverse and longitudinal frequencies of the $j^{th}$ polar phonon mode, respectively, and $\gamma_{TO}$ and $\gamma_{LO}$ denote their corresponding damping constants.

The oscillator strength $\Delta\varepsilon_j$ [i.e., contribution of the phonon mode to the static dielectric constant $\varepsilon(0)$] of the $j^{\text{th}}$ polar phonon can be calculated from the formula

$$\Delta\varepsilon_j = \frac{\varepsilon_\infty}{\omega_{TO_j}^2} \frac{\prod_k (\omega_{LO_k}^2 - \omega_{TO_j}^2)}{\prod_{k\neq j} (\omega_{TO_k}^2 - \omega_{TO_j}^2)} \qquad (2)$$

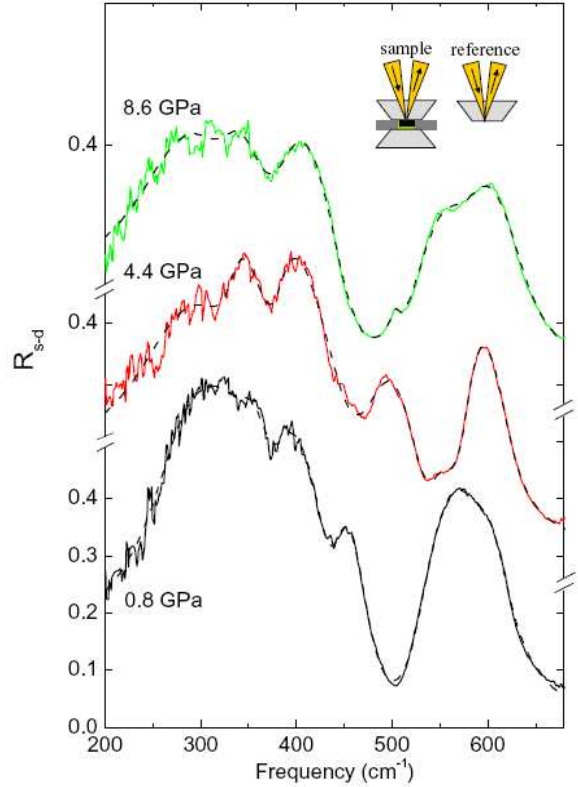

**Figure 5** (Color online)
Room-temperature reflectivity $R_{s-d}$ spectra of BiFeO$_3$ for three selected pressures (0.8, 4.4, 8.6 GPa); the spectra are offset along the vertical axis for clarity. The dashed lines are the fits with the generalized oscillator model according to Eq. (1) (see text for details). Inset: Measurement geometry for the reflectivity measurements, as described in the text.

The four-parameter oscillator model [Eq. (1)] follows from the general properties of the dielectric function in a polarizable lattice (pole at transverse and zero at longitudinal eigenfrequencies of polar phonons) and it is able to describe the permittivity of dielectrics in most cases. However, it has a drawback since a certain combination of parameter values in Eq. (1) may result in unphysical values of the complex permittivity[76, 77] (for example, negative losses or finite conductivity at infinite frequency). Therefore, in our fitting procedure of the infrared reflectivity we restricted the parameter values to those that result in an optical conductivity vanishing at frequencies much higher than the phonon eigen-frequencies.

The dielectric function $\varepsilon(\omega)$ [Eq. (1)] is directly related to the measured reflectivity $R_{s-d}(\omega)$ at the sample-diamond interface by the Fresnel equation:

$$R_{s-d}(\omega) = \left| \frac{\sqrt{\varepsilon(\omega)} - n_{dia}}{\sqrt{\varepsilon(\omega)} + n_{dia}} \right|^2 \quad (3)$$

The pressure-dependence of the high-frequency permittivity $\varepsilon_\infty$ used in our fitting was calculated according to the Clausius-Mossotti relation:

$$\frac{\varepsilon_\infty(P) - 1}{\varepsilon_\infty(P) + 2} = \frac{\alpha}{3\varepsilon_0 V(P)}, \quad (4)$$

where α is the electronic polarizability of the unit cell, which was obtained from the lowest-pressure data. The high-frequency permittivity $\varepsilon_\infty$ as a function of pressure, calculated with Eq. (4) using the experimentally determined pressure dependence of the unit cell volume shows nearly linear increase with pressure coefficient of 0.16 GPa$^{-1}$. The estimated value of $\varepsilon_\infty$ at ambient pressure is 6.8. It is higher than the value of 4.0 reported for BiFeO$_3$ ceramics[75], however, lower than $\varepsilon_\infty$ = 9.0 reported for single crystals.[52] Therefore, the $\varepsilon_\infty$ value used in this work is reasonable. However, its precision is critically dependent on several parameters that practically uncontrollable in pressure experiments (like surface quality, parasitic reflections from diamond anvil interfaces etc.).

The reflectivity spectra could be well-fitted with the generalized-oscillator model according to Eq. (1). As examples, we show in Fig. 5 reflectivity spectra $R_{s-d}$ of BiFeO$_3$ at three selected pressures and the corresponding fits with the generalized-oscillator model. Below $P_{c1}$ = 3 GPa the reflectivity spectra in the measured frequency range can be well-fitted using 6 oscillator terms. Above 3 GPa an additional oscillator term is needed for a reasonable fit of the spectra. Finally, above 7.5 GPa the number of oscillators reduces to six again. The pressure dependence of the transverse phonon frequencies is shown in Fig. 6.

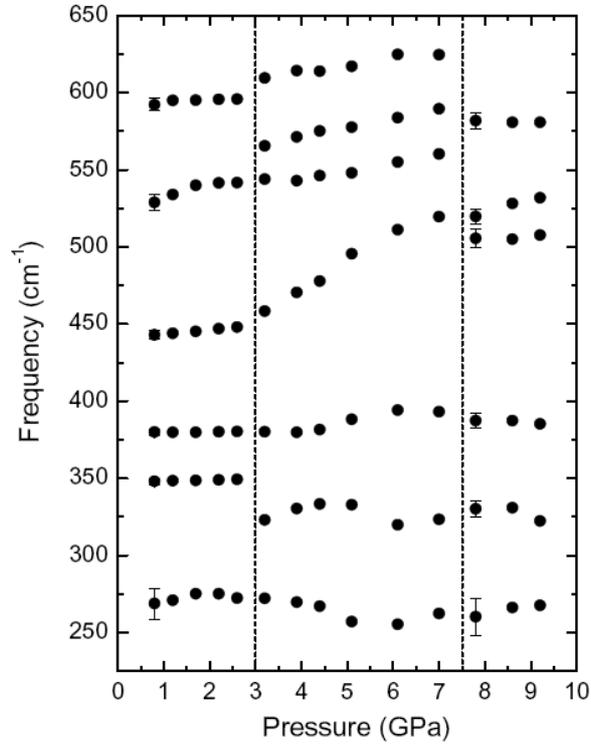

**Figure 6**
Frequencies of the transverse optical phonons in BiFeO$_3$ as a function of pressure, obtained by fitting the reflectivity spectra $R_{s-d}(\omega)$ with the generalized-oscillator model. The vertical dashed lines indicate the pressures of the two phase transitions.

The factor-group analysis predicts 13 infrared- and Raman-active phonon modes for the room temperature $R3c$ phase of BiFeO$_3$. They can be classified according to the irreducible representations $4A_1 + 9E$, i.e., there are 4 $A_1$ modes polarized along the direction of the spontaneous polarization and 9 $E$ doublets polarized normal to this direction. In addition, there are 5 $A_2$ silent modes. The frequencies of the optical phonons have been calculated theoretically[78] and determined experimentally by infrared[52] and Raman[79, 80] spectroscopy on single BiFeO$_3$ crystals. According to the fit of our data with the generalized-oscillator model the transverse optical modes are located at 269, 348, 380, 443, 529 and 592 cm$^{-1}$ for the lowest measured pressure (0.8 GPa).

| $\omega_{TO}(\gamma_{TO})$ | $\omega_{TO}^{amb}(\gamma_{TO}^{amb})$ | $\omega_{LO}(\gamma_{LO})$ | $\Delta\epsilon$ | $\Delta\epsilon^{amb}$ |
|---|---|---|---|---|
| 269 (51) | 262 (9.1) | 348 (41) | 18.2 | 14.8 |
|  | 274 (33.5) |  |  | 2.45 |
| 348 (36) | 340 (17.4) | 374 (43) | 0.023 | 0.27 |
| 380 (41) | 375 (21.6) | 433 (43) | 0.32 | 0.475 |
| 443 (33) | 433 (33.8) | 472 (44) | 0.15 | 0.301 |
| 529 (48) | 521 (41.3) | 588 (48) | 0.69 | 1.14 |
| 592 (46) |  | 614 (37) | 0.019 |  |

**Table 4**
Room-temperature fitting parameters from Eq. (1) to describe the reflectivity spectrum of BiFeO3 at 0.8 GPa, compared to the room-temperature parameters obtained at ambient pressure by Lobo et al.[52], denoted by $\omega_{TO}^{amb}$, $\gamma_{TO}^{amb}$ and $\Delta\varepsilon^{amb}$

In Table 4 we list the frequencies of the transverse and longitudinal optical modes obtained by our infrared reflectivity measurements on single crystals at the lowest pressure together with the ambient-pressure results for a BiFeO$_3$ single crystal obtained by Lobo *et al.*[52] There is a very good agreement between the transverse phonon frequencies $\omega_{TO}$ obtained from our fit and $\omega_{TO}^{amb}$ from ref.[52]. However, the damping constants $\gamma_{TO}$ are higher in the case of our pressure measurements. The difference in the far-infrared reflectivity spectra $R_{s-d}(\omega)$ for the two sets of parameters given in Table 4 is illustrated in Fig. 5. Obviously, both reflectivity spectra look similar and differ only in the overall reflectivity level and the sharpness of the phonon dips. Such broadening of the phonon modes under high pressure is rather common especially in the case of a solid pressure transmitting medium.[81] The mode at 274 cm$^{-1}$, which produces a small dip in the reflectivity curve (marked by an asterisk in Fig. 7) observed by Lobo *et al.*[52], becomes weaker due to the broadening effect in our pressure measurements. It could not therefore be resolved reliably in the measured spectra and was therefore neglected in our fitting procedure.

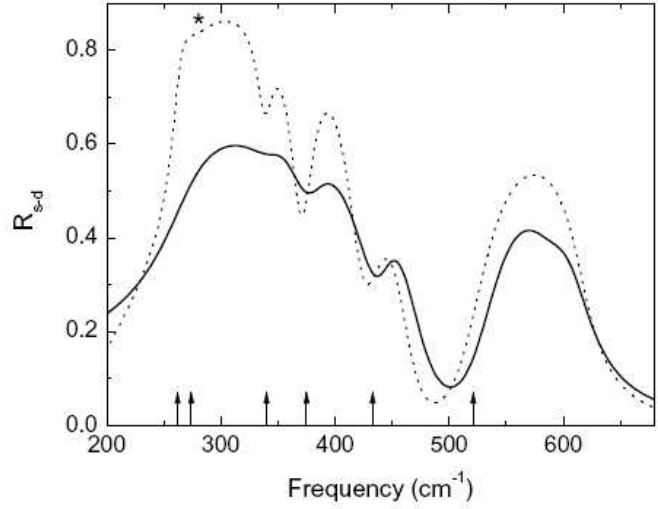

**Figure 7**
Fit of the measured reflectivity spectrum of $BiFeO_3$ at 0.8 GPa (solid line) compared to the simulated ambient pressure spectrum in the diamond anvil cell using the fitting parameters from Ref. 28 (dashed line). The arrows indicate the frequencies of TO phonons found by Lobo et al. [52]. The asterisk marks the kink produced by the mode at 274 cm−1.

All the phonon modes listed in Table 4, besides the weak mode at 592 cm$^{-1}$, belong to the $E$ representation; i.e., they are aligned perpendicular to the direction of spontaneous polarization $[111]_{pc}$. This indicates that the electric field of the synchrotron radiation used in our experiment was polarized approximately along the $[-110]_{pc}$ direction, similar to the experiment of Lobo *et al.*[52]

The evolution of the optical conductivity $\sigma'(\omega) = \omega\varepsilon_0\varepsilon''(\omega)$ with increase of pressure is shown in Fig. 8. We can see the drastic changes of the optical conductivity spectra across the transition pressures $p_{c1}$ = 3 GPa and $p_{c2}$ = 7.5 GPa.

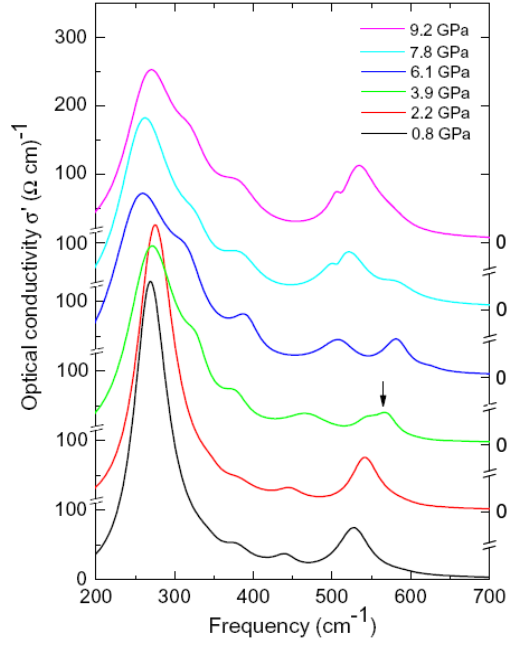

**Figure 8** (Color online)
Real part σ′(ω) of the optical conductivity of $BiFeO_3$ for selected pressures, obtained by fitting the reflectivity spectra $R_{s-d}(\omega)$ with the generalized oscillator model; the spectra are offset along the vertical axis for clarity. The arrow marks the position of the phonon mode at 565 $cm^{-1}$ emerging above 3 GPa.

The five detected phonon modes can be assigned to the bending and stretching modes of the $FeO_6$ octahedra, which exhibit a displacement of the $Fe^{3+}$ cations from their centrosymmetric position along the pseudo-cubic $[111]_{pc}$ direction.[37, 38] The change in the pressure dependence of the phonon mode frequencies at $p_{c1}$ and $p_{c2}$ could thus be assigned to changes in the octahedral distortion. The Bi ions are mainly involved in the lower-frequency (<200 $cm^{-1}$) modes located below the measured frequency-range of this study.

Our pressure-dependent far-infrared data confirm the occurrence of two phase transitions in $BiFeO_3$. The most significant spectral signature of the phase transition at 3 GPa is the appearance of a phonon mode at 565 $cm^{-1}$ (see Figs. 5 and 8). Furthermore, the pressure-dependence of the frequency of the other TO phonon modes demonstrates anomalies across the transition pressure (change of the slope of the frequency shift). Complementary to this finding, the Raman measurements under pressure detected the appearance of new modes and clear anomalies around 3 GPa only for the modes below 250 cm$^{-1}$ which were not accessible by our far-infrared study.

According to our XRD data a second transition into a paraelectric phase with *Pnma* symmetry occurs. Since the unit cell of the orthorhombic perovskite with *Pnma* space group contains 4 formula units, i.e., twice more atoms than the rhombohedral *R*3*c* unit cell of the BiFeO$_3$, the number of the phonon modes should be doubled in the paraelectric phase. In analogy with the perovskite LaMnO$_3$, there should be in total 25 infrared modes 9B$_{1u}$ +7B$_{2u}$ +9B$_{3u}$ in the paraelectric phase of BiFeO$_3$. The increased number of modes in the *Pnma* phase compared to 13 modes in the *R*3*c* phase should originate from the splitting of the *E* symmetry doublets and the general doubling of all modes due to the unit cell doubling. We would therefore expect to observe a splitting of the phonon modes across the transition pressure, although some modes can vanish due to the selection rules. Such effects were reported in pressure-dependent Raman measurements of BiFeO$_3$ crystals around 9-10 GPa.[39] Our infrared measurements demonstrate a similar effect: above 7.5 GPa the mode located at 520 cm$^{-1}$ below the transition pressure splits into two modes (see Figs. 5, 8). Thus, our infrared study confirms the second pressure-induced phase transition but the transition pressure $p_{c2} \approx 7.5$ GPa is somewhat lower than the value of 9-10 GPa observed by XRD and previous[39] Raman studies. This difference in pressure can be understood by the different pressure transmitting media used in the two experimental investigations (cryogenic liquids in the Raman[39] and present x-ray measurements and solid CsI in the case of IR spectroscopy), since under more hydrostatic conditions the transition is expected to occur at higher pressure.[82]

### IV.    Concluding remarks

In summary, our pressure-dependent IR and X-ray scattering study reveals that BFO presents significant pressure-instabilities in agreement with recent theoretical predictions.[40] A first structural phase transition occurs as low as 3 GPa towards a distorted monoclinic perovskite structure which is characterized by the superimposition of tilts *and* cation displacements. With further increasing pressure the cation displacements of $BiFeO_3$ are reduced and finally suppressed around 10 GPa leading to the non-polar *Pnma* structure in agreement with recent[40] theoretical ab-initio predictions (that have not predicted the occurrence of the intermediate phase). Contrary to earlier experimental[41-45] and theoretical investigations[46] of BFO where no structural phase transition was reported, our study provides evidence that BFO presents further structural instabilities below 15 GPa (added note: a very recent[67], yet unpublished, work provides further experimental evidence for this)

It appears that a complex competition between the oxygen octahedra tilting and the polar character especially through the Bi lone pair electron conditions the intermediate monoclinic phase, which we believe to be a general feature for Bi-based perovskite compounds.

**Acknowledgements**

We acknowledge the ANKA Angströmquelle Karlsruhe and the European Synchrotron Radiation Facility (ESRF) for the provision of beam time. Specifically, we would like to thank B. Gasharova, Y.-L. Mathis, D. Moss, and M. Süpfle for assistance using the beamline ANKA-IR and M. Hanfland at the ESRF for beamline support. Financial support by the Bayerische Forschungsstiftung and the DFG through the SFB 484 is gratefully acknowledged.

Support from the French National Research Agency (ANR Blanc) is greatly acknowledged by P.B, R.H., B.D. & JK. Furthermore, JK thanks the European network of excellence FAME and the European STREP MaCoMuFi for financial support.

Finally, the authors thank J.F. Scott and L. Bellaiche for fruitful discussions.